# Thermal conductivity and enhanced thermoelectric performance of SnTe bilayer

*Abhiyan Pandit,[a,]\* Raad Haleoot,[b] and Bothina Hamad[a,c]*

**ABSTRACT**

Tin chalcogenides (SnS, SnSe and SnTe) are found to have improved thermoelectric properties upon the reduction of their dimensionality. Here we found the tilted AA + s stacked two-dimensional (2D) SnTe bilayer as the most stable phase among several stackings as predicted by the structural optimization and phonon transport properties. The carrier mobility and relaxation time are evaluated using the deformation potential theory – these are found to be relatively high due to the high 2D elastic modulus, low deformation potential constant and moderate effective masses. The SnTe bilayer shows high Seebeck coefficient (> 400 µV/K), high electrical conductivity and ultralow lattice thermal conductivity (< 1.91 Wm$^{-1}$K$^{-1}$). High TE figure of merit (ZT) values, as high as 4.61 along zigzag direction, are predicted for SnTe bilayer within the carrier concentration range of the order 10$^{12}$ - 10$^{13}$ cm$^{-2}$. These ZT values are much enhanced as compared to the bulk as well as monolayer SnTe and other 2D compounds.

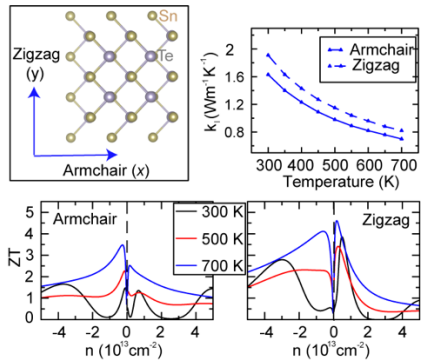

**Figure:** Graphical Abstract

## 1. INTRODUCTION

Thermoelectric (TE) materials are solid state semiconductors that can be used for harvesting the waste heat produced in thermal power generation and convert it into electricity or vice versa. They are considered to be one of the promising sources of renewable energy. The efficiency of a TE device can be evaluated by using the dimensionless figure of merit (*ZT*) given as:

$$ZT = \frac{S^2 \sigma T}{(\kappa_l + \kappa_e)}, \qquad (1)$$

where S, $\sigma$ and T represent Seebeck coefficient, electrical conductivity and temperature, respectively; $\kappa_l$ and $\kappa_e$ represent the lattice thermal and electronic thermal conductivities. Group IV-VI compounds are widely used in photovoltaics and thermoelectrics as they are earth-abundant, less toxic, chemically stable and environmentally compatible.[1–6] Germanium and tin chalcogenides (GeS, GeSe, SnS and SnSe) have attracted a great deal of attention due to their large Seebeck coefficient, high power factor and low thermal conductivity.[7] In addition, lead and tin chalcogenides are investigated intensively in the TE studies due to their intrinsic lattice anharmonicity and structural anisotropy, which is useful in improving the TE performance.[3,8–10]

Lowering the dimensionality has been one of the efficient methods on increasing the TE efficiency due to the increase of the Seebeck coefficient and the electrical conductivity, and reduction of the lattice thermal conductivity.[11–13] Two-dimensional (2D) monolayers (MLs) of black phosphorus, and group-IV chalcogenides are found to have much enhanced TE performance than their bulk forms.[14–16] Moreover, diverse structural, electronic and ferroelectric properties along with their stability have been investigated in ultrathin SnTe layers.[17,18] The SnTe ML has zigzag and armchair like projections of atoms within the plane. Bilayer structure of a layered compound can be created by stacking two MLs (one at the top of other in a certain pattern). Importantly, the most stable type of stacking (AA, AB etc.) at room temperature entirely depends on the chemical and structural property of the corresponding compound – it can be tested through structural optimization, phonon dispersion relation and so on. The AA stacked bilayer structure of group IV-VI compounds, such as: GeSe and SnSe, was found to be the most stable.[19,20] Recent studies on SnTe MLs predicted high ZT values over 3.81 (in $\beta'$ (hexagonal)-phase at 900 K) and ~ 1.46 (in $\gamma$ (rectangular)-phase at 700 K),[16,21] which indicates that SnTe MLs are promising candidates for TE applications. Although there are intensive studies on the TE properties of SnTe MLs, reports on the 2D bilayer structures are scarce, which is the stimulus of the present work.

In this paper, the structural, electronic, lattice dynamics and thermoelectric properties of AA + s (upper ML shifted by $a_1/6$) SnTe bilayer are presented. The rest of this paper is organized as: section 2 discusses the computational methods, section 3 includes the results and discussion, and finally the concluding remarks are presented in section 4.

## 2. COMPUTATIONAL METHODOLOGY

Calculations are based on the density functional theory combined with the Boltzmann transport formalism as implemented in the VASP package.[22] The exchange and correlation energy contributions were described by the Perdew-Burke-Ernzerhof (PBE) functional,[23] and the ionic cores of the atoms were represented within the projector augmented-wave (PAW) formalisms as the pseudopotentials.[24] The van der Waals interaction is included by using vdW-DF scheme.[25] The optimized cut-off energy of the plane wave expansion is set to 500 eV. The Brillouin zone (BZ) is sampled using 18×18×1 Monkhorst-Pack (MP) $k$-point mesh, centered at Γ-point for the structural optimization. A vacuum region of 15 Å perpendicular to the $z$-direction is employed to avoid the interactions between the periodic images of the unit cell. Total energies and forces were converged up to 0.001 meV per atom and 0.001 eV/Å, respectively.

The electronic transport calculations were performed using the semiclassical Boltzmann transport theory as implemented in BoltzTraP2 package (version - 20.2.1). A dense $k$-point mesh of 36×36×1 is used to compute the density of states calculations. The deformation potential (DP) theory[26] is applied to study carrier mobility and the relaxation time as the DP theory has been extensively applied in similar 2D and 1D materials.[15,20,27–30]

The lattice dynamics and vibrational stability were investigated using PHONOPY.[31] A supercell of 4×6×1 of SnTe bilayer unit cell with the 4×4×1 $k$-point mesh is used for the phonon dispersion and the second-order (harmonic) interatomic force constants (IFCs) calculation with a default finite displacement of 0.01 Å. The third-order (anharmonic) IFCs are calculated using a 4×4×1 unit cell with the 4×4×1 $k$-point mesh, where the interactions are considered up to the fifth nearest neighbors. The second- and third-order IFCs are evaluated as $\phi_{ij}^{\alpha\beta} = \frac{\partial^2 V}{\partial r_\alpha^i r_\beta^j}$ and $\phi_{ijk}^{\alpha\beta\gamma} = \frac{\partial^3 V}{\partial r_\alpha^i r_\beta^j r_\gamma^k}$, respectively; where $V$ is the potential energy of the phonon system, $r$ represents the atomic positions of the corresponding atoms ($i, j$ and $k$) and $\alpha, \beta, \gamma$ represent the cartesian directions of the finite displacements.[32,33] The calculated second- and third-order IFCs are used in ShengBTE code[32] to evaluate the lattice thermal conductivity ($k_l$) and other thermodynamic parameters. The CONTROL file with the input tags used in this calculation is available in the supporting information.

## 3. RESULTS AND DISCUSSIONS

### 3.1. Structural and electronic properties

The structural optimization was carried out by taking SnTe ML as a basic building block with various-stacking configurations. The initial configuration was the AA-stacking, and then the upper ML is shifted by some fractions of the lattice constant along $x$-direction ($a_1$). The cohesive energy here can be calculated as:

$$E_c = E_{(SnTe\text{-}bilayer)} - m \times E_{(Sn)} - n \times E_{(Te)}, \quad (2)$$

where $E_{(SnTe\text{-}bilayer)}$, $E_{(Sn)}$ and $E_{(Te)}$ are the total energies of SnTe bilayer, Sn and Te, respectively; $m$ and $n$ are the numbers of Sn and Te in the SnTe bilayer. As the SnTe ML has 4 atoms, the unit cell of SnTe bilayer is composed of 8 atoms (4 Sn and 4 Te atoms). The structural relaxation and cohesive energy calculations predict that the AA + $\frac{1}{6}$ $a_1$ stacked SnTe bilayer, where $a_1$ is the lattice constant along armchair ($x$-) direction (AA + s hereafter with s used for the shift), is the most stable structure with the lowest cohesive energy of -20.27 eV among the various stackings (AA, AA + $\frac{1}{2}$ $a_1$, etc.). The optimized lattice constants are 4.63 and 4.56 Å along armchair and zigzag directions. The optimized structure of AA + s stacked SnTe bilayer is shown in Figure 1, where the Figure 1a represents the top view along with clearly distinctive armchair ($x$-) and zigzag ($y$-) directions. Figures 1b and 1c are the side views perpendicular to the zigzag and armchair directions, respectively. It can be clearly noticed that the SnTe bilayer is tilted/distorted (with an angle of 77º) along $z$-$x$ plane as shown in Figure 1b. The detailed optimized lattice constants along with the atomic positions are presented in the supporting information.

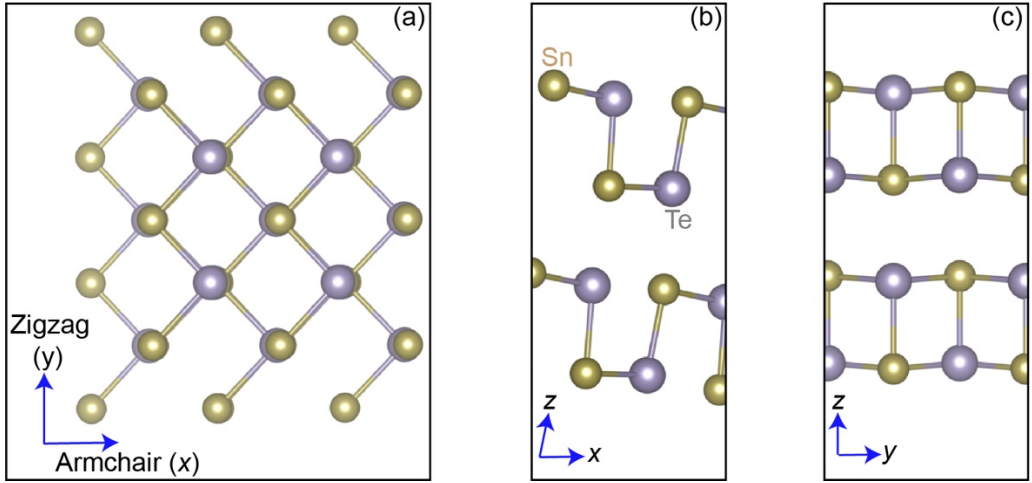

**Figure 1.** Crystal structure of AA + s stacked SnTe bilayer: (a) top view, (b) side view perpendicular to the zigzag direction and (c) side view perpendicular to the armchair direction.

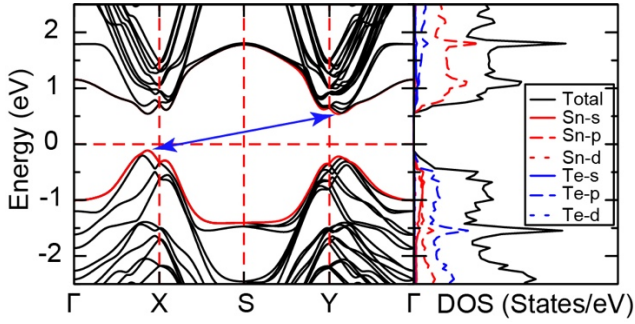

**Figure 2.** Electronic band structure and DOS of AA + s stacked SnTe bilayer. The electronic bands with red color indicate the CBM and VBM.

The electronic band structure of AA + s stacked SnTe bilayer is presented in Figure 2. The SnTe bilayer is found to have an indirect band gap of 0.72 eV without spin orbit coupling (SOC) interaction, which reduces to 0.65 eV upon including the SOC. The conduction band minima (CBM) is found along Y – Γ path and valence band maxima (VBM) is along Γ – X path as shown in Figure 2. The electronic band gap here is in agreement with the previous theoretical result of ~ 0.63 eV,[34] and smaller than the experimental value ~ 0.9 eV.[17] The difference in between the calculated and experimental result is due to the typical problem of Kohn Sham theory in underestimating the band gaps up to 50%.[35,36] The electronic density of states (DOS) plot shows that *p*-orbitals of Sn and Te atoms have a significant contribution to the sharper peaks at the conduction (*n*-type) and valence (*p*-type) bands, respectively. As the peaks near the valence and conduction band are nearly equal, AA + s stacked SnTe bilayer can be suitable for both *p*- or *n*-type dopings in TE applications.

### 3.2. Carrier mobility and relaxation time

The carrier mobility ($\mu$) for a 2D system is evaluated using the following formula:[28,37]

$$\mu_{2D} = \frac{2e\hbar^3 C^{2D}}{3k_B T m^* m_d E_1^2} \tag{3}$$

where $e$ is the electronic charge, $\hbar$ is the reduced Planck constant, $T$ represents the temperature, $C^{2D}$ is the 2D elastic modulus, $E_1$ is the DP constant, $m^*$ is the effective mass and $m_d$ is defined as $m_d = \sqrt{m_x m_y}$. Here $m_x$ and $m_y$ are the effective masses along armchair (*x*-) and zigzag (*y*-) directions. The value of $C^{2D}$ can be obtained by fitting the energy-strain curves along armchair/zigzag directions. The two curves are nearly identical as shown in Figure S1 (supporting information), due to which the values of $C^{2D}$ along the two directions are very close (86.65 N/m along armchair and 89.84 N/m along zigzag direction). The DP constant ($E_1$) can be defined as $E_1 = \frac{\partial E_{edge}}{\partial \delta}$, equivalent to the straight line fitting, where $E_{edge}$ is the energy of CBM or VBM and $\delta$ is the uniaxial strain. The shift in CBM by applying the uniaxial strain along both directions is shown in Figure S2 (supporting information). The effective mass ($m^*$) for charge transport can be computed from the band structure. The $m^*$ value of the electron (hole) at CBM (VBM) along armchair and zigzag directions can be calculated as $m^* = \frac{\hbar^2}{\frac{\partial^2 \varepsilon(k)}{\partial k_\alpha \partial k_\beta}}$, where the term $\frac{\partial^2 \varepsilon(k)}{\partial k_\alpha \partial k_\beta}$ is obtained as the second derivative of the band energy $E(k)$ with respect to $k$ vector (along Γ – X path for armchair and Γ – Y path for zigzag direction). At 300 K, a similar effective masses for electrons (0.15$m_e$ along *x*- and *y*-directions) and holes (0.12$m_e$ along *x*-direction and 0.15 $m_e$ along *y*-direction) are predicted, which is attributed to the similar parabolic nature of band dispersions of CBM and VBM (as in Figure 2). On the basis of the obtained values of $C^{2D}$, $E_1$ and $m^*$, the carrier mobility ($\mu_{2D}$) can be calculated using Eq. (3). Normally adopted value of relaxation time ($\tau$) is in the order of 10$^{-14}$ s within the constant relaxation time approximation,[7,38–40] but the actual value of $\tau$ of a specific material may vary depending upon its physical property.[15,20,21,28] Here the relaxation time ($\tau$) is evaluated using the relation:

$$\tau = \frac{m^* \mu_{2D}}{e} \tag{4}$$

The calculated $C^{2D}$, $E_l$, $m^*$, $\mu_{2D}$ and $\tau$ for AA + s stacked SnTe bilayer at 300 K and higher temperatures are compiled in Table 1. The SnTe bilayer is found to have relatively high relaxation time and mobility as compared to other group IV-VI layered compounds as a consequence of low deformation potential, high elastic modulus and moderate values of effective mass.[15,20,21]

**Table 1.** Deformation potential constant ($E_l$), 2D elastic constant ($C^{2D}$), effective mass ($m^*$), carrier mobility ($\mu_{2D}$) and relaxation time ($\tau$) of SnTe bilayer at 300K, 500 K and 700 K, in armchair and zigzag directions.

| Direction | Carrier | $E_l$ (eV) | $C^{2D}$ (N/m) | $m^*$ ($m_e$) | $\mu_{2D}$ (cm$^2$V$^{-1}$s$^{-1}$) | $\tau$ ($10^{-14}$ s) 300 K | 500 K | 700 K |
|---|---|---|---|---|---|---|---|---|
| Armchair | e | - 4.09 | 86.65 | 0.15 | 3158.06 | 27.47 | 16.48 | 11.77 |
|  | h | - 4.09 | 86.65 | 0.12 | 4413.52 | 30.71 | 18.43 | 13.16 |
| Zigzag | e | 4.68 | 89.84 | 0.15 | 2503.24 | 21.77 | 13.06 | 9.33 |
|  | h | 4.68 | 89.84 | 0.15 | 2798.70 | 24.34 | 14.61 | 10.43 |

### 3.3. Lattice dynamics and thermal conductivity

The TE efficiency of a material is contributed by lattice and electronic transport properties. The phonon dispersion relation is a good measure to test the vibrational stability and lattice dynamics of a system. The positive phonon frequencies obtained from Figure 3a indicate that AA + s stacked SnTe bilayer is dynamically stable. As the unit cell has eight atoms there are 24 (3N) phonon branches, where the lowest three are the acoustic modes and rest nine are the optical modes. Out of three acoustic modes, two are transverse acoustic (TA) and one is longitudinal acoustic (LA). TA1 and TA2 are clearly noticed to be degenerate along the X – S – Y path as in Figure 3a. The phonon DOS is depicted in Figure 3b – it shows that both Sn and Te atoms occupy the high/low frequency region in the way since there is no significant difference in atomic weight of both atoms. The group velocity ($v_g$), which is defined as $v_g = \omega(k)/\partial k$, is an important factor in determining the $\kappa_l$ as $\kappa_l$ is directly proportional to $v_g$ (see Eq. 5). From Figure 3c, the minority of the TA2, LA and optical branches are found to contribute to higher group velocity.

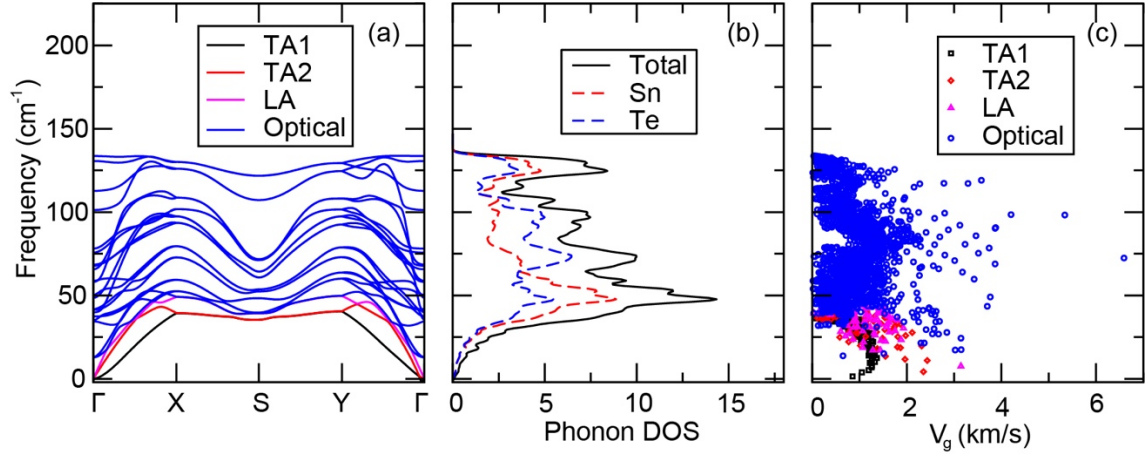

**Figure 3.** (a) Phonon dispersion relation, (b) Phonon DOS and (c) Group velocity of AA + s stacked SnTe bilayer.

In addition to $v_g$, Grüneisen parameter ($\gamma$) and the specific heat capacity ($C_v$) are also important quantities to calculate $\kappa_l$. The Grüneisen parameter ($\gamma$), which measures the anharmonicity of a system, is inversely related to $k_l$ according to Slack's theory.[41] Accordingly, the value of $\gamma$ is found to increase with increasing the temperature and saturates at T = 700 K as shown in Figure 4a. The temperature dependence of $C_v$, which is obtained using the second and third order IFCs are depicted in Figure 4b. The value of $C_v$ is found to increase by increasing temperature and starts to saturate near 700 K approaching the classical limit of Dulong and Petit[42]. The total converged scattering rate along the irreducible $q$-points at 300 K is depicted in Figure 4c, which indicates that the scattering rate is larger for the branches that have the larger value of $v_g$. The effect of grain size on the $\kappa_l$ values at different temperatures (300 K, 500 K and 700 K) can be investigated by plotting the cumulative $\kappa_l$ as function of mean free path (MFP) (see Figure 4d). As an example, at 300 K, a MFP of 0.79 Å is required to change $\kappa_l$ value by 50% (from 0.165 to 0.252 Wm$^{-1}$K$^{-1}$). Large MFP phonons contribute to a significant amount on the cumulative $k_l$ values at a smaller temperature (300 K). The cumulative $\kappa_l$ values are found to increase with increasing the MFP (for all temperatures), and saturate at a certain threshold.

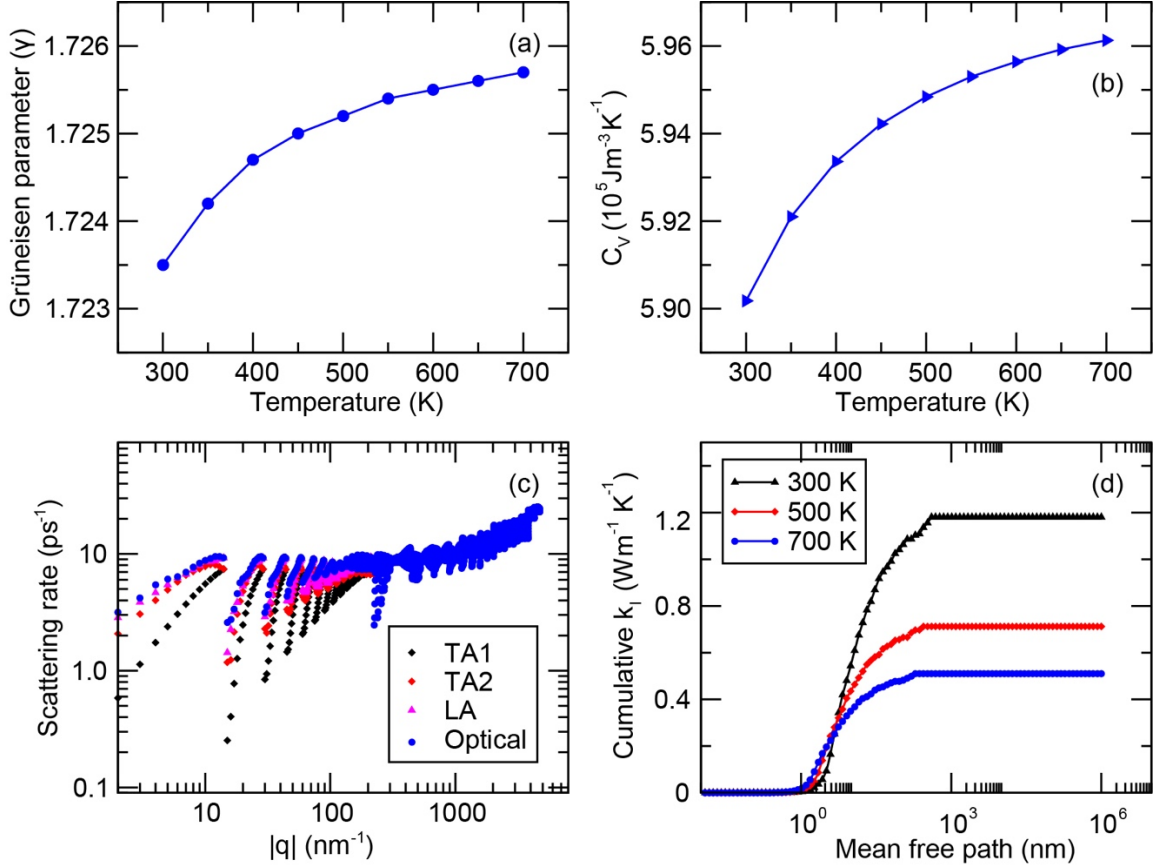

**Figure 4.** (a) Grüneisen parameter and (b) specific heat capacity as a function of temperature. (c) Phonon scattering rate at 300 K for different phonon branches and (d) the cumulative $k_l$ at 300 K, 500 K and 700 K as a function of mean free path.

The lattice thermal conductivity ($\kappa_l$) is calculated using the equation:[32]

$$k_l^{\alpha\beta} = \frac{1}{k_B T^2 \Omega N} \sum_\lambda f_0(f_0 + 1)(\hbar w_\lambda)^2 v_\lambda^\alpha F_\lambda^\beta, \qquad (5)$$

where $k_B$, $N$, $T$ and $f_0$ are the Boltzmann constant, number of $q$-points, temperature and the phonon (Bose-Einstein) distribution function, respectively; $\omega_\lambda$ and $v_\lambda$ are the angular frequency and group velocity of phonon mode $\lambda$, respectively. Using the iterative method for phonons, the calculated $k_l$ values of AA + s stacked SnTe bilayer along armchair and zigzag directions as a function of temperature is shown in Figure 5, which shows an anisotropic behavior. This anisotropy in $k_l$ is closely related to the anisotropy of phonon group velocity as depicted mathematically in Eq. (5).[43,44] The calculated value of $k_l$ is 1.63 (1.91) Wm$^{-1}$K$^{-1}$ at 300 K along armchair (zigzag) direction, which is much lower as compared to other 2D compounds such as graphene,[45] phosphorene,[46] SnS, SnSe MLs;[15] higher as compared to SnSe bilayer,[20] and slightly higher than γ-SnTe MLs.[16] The value of $k_l$ decreases with the increase in temperature – this phenomena is due to the anharmonic phonon scattering resulting from the lattice vibration at higher temperatures.[47] The lower $k_l$ value contributes to the enhancement of the TE performance of a material by the increase of its figure of merit (*ZT*).

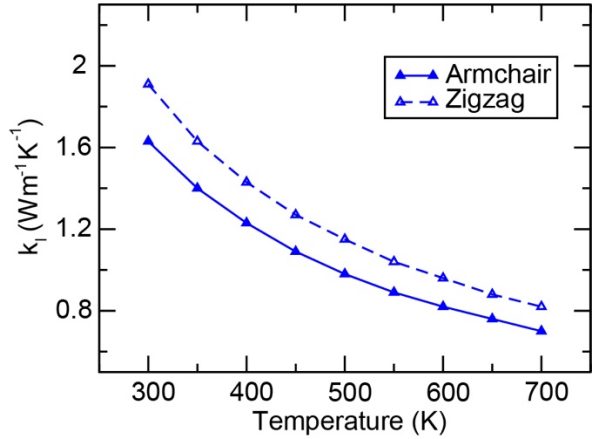

**Figure 5.** The lattice thermal conductivity ($k_l$) of AA + s stacked SnTe bilayer as a function of temperature along armchair and zigzag directions.

### 3.4. Electronic transport properties and *ZT*

The Seebeck coefficient (*S*) for the metals or degenerate semiconductors can be calculated using the relation:[12]

$$S = \frac{8\pi^2 k_B^2}{3eh^2} m^* T \left(\frac{\pi}{3n}\right)^{2/3}, \tag{6}$$

where $m^*$ is the effective mass of the carrier and *n* is the carrier concentration. On the other hand, the electrical conductivity (σ) is related to *n* and $m^*$ as:

$$\sigma = \frac{ne^2\tau}{m^*}. \tag{7}$$

A similar carrier concentration range (of the order $10^{12} - 10^{13}$ cm$^{-2}$) is adopted for a better comparison with similar layered compounds.[15,20,21] The Seebeck coefficient (*S*) as a function of carrier concentration (*n*) for the SnTe bilayer along armchair and zigzag directions at 300 K, 500 K and 700 K is presented in Figure 6a, where negative values of *n* represent the electron (*n*-type) and positive values represent the hole (*p*-type) concentration. One can notice an abrupt increase in the values of |*S*| near *n* = 0. Further, the |*S*| values decrease by increasing the carrier concentration in agreement with Eq. (6) with a discontinuity at *n* = 0. The Seebeck coefficients along armchair and zigzag directions are found to have similar values as a result of their similar effective masses.[12,21] In addition, the value of S increases with increasing temperature – this implies the occurrence of more frequent carrier scattering events at higher temperatures (Eq. (6)), in agreement with the results of bulk SnTe.[48,49] The value of S for SnTe bilayer here is significantly larger than that of bulk SnTe,[48] but smaller as compared to the SnTe MLs.[16,21]

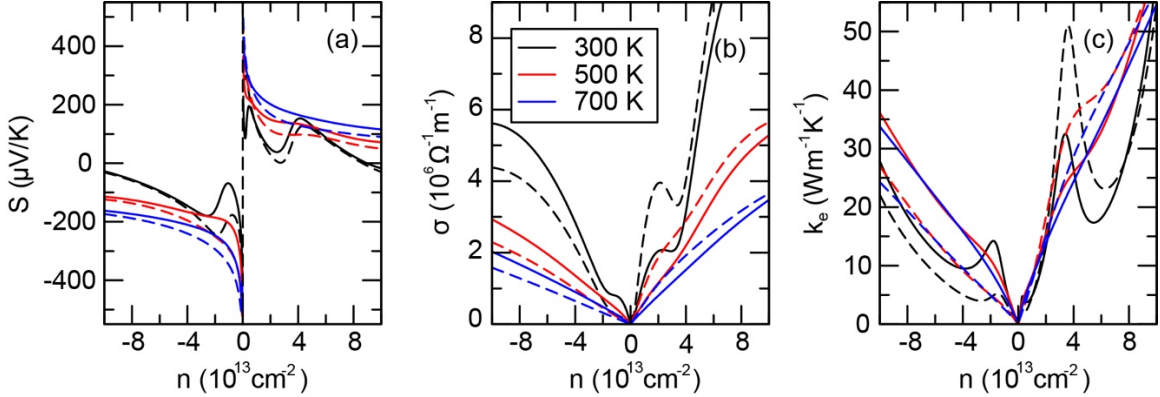

**Figure 6.** Electrical transport properties of SnTe bilayer as a function of carrier concentration (n) at 300 K, 500 K and 700 K. (a) Seebeck coefficient (S), (b) electrical conductivity ($\sigma$) and (c) electronic thermal conductivity ($k_e$). The positive and negative values of $n$ indicate holes and electrons concentration respectively. Curves with solid lines represent armchair direction and those with dotted lines represent the zigzag direction.

The electrical conductivity ($\sigma$) and electronic thermal conductivity ($\kappa_e$) are obtained from the BoltzTraP2 code[50] as $\sigma/\tau$ and $\kappa_e/\tau$ within the constant time approximation. The computed values of $\tau$ (see Table I) are used to obtain $\sigma$ and $\kappa_e$ at different temperatures. Figure 6b shows the calculated $\sigma$ as a function of carrier concentration ($n$) at 300 K, 500 K and 700 K along armchair and zigzag directions. The value of $\sigma$ increases with the increase of $n$ as they are directly proportional to each other (Eq. (7)). The electrical conductivity ($\sigma$) is found to decrease on increasing the temperature – this is related to the intrinsic electrons scattering mechanism at higher temperatures. The electrical conductivity for SnTe bilayer here is lower as compared to its bulk form due to the increase in band gap. Interestingly, $\sigma$ values are slightly higher than those of SnTe ML,[16,21,51] which suggests a possible enhancement in the TE performance of the bilayer structure. The electronic thermal conductivity ($\kappa_e$) comes from the electrons and holes transporting heat in a system. The calculated $\kappa_e$ as a function of $n$ at different temperatures is depicted in Figure 6c, where $\kappa_e$ increases with increasing $n$. The electronic thermal conductivity is found to be relatively larger for p-type (hole) charge carriers than n-type (electron) charge carriers. There is an increase in the value of $\kappa_e$ by increasing temperature because of the excitation of more electrons at higher temperatures.

The values of power factor ($S^2\sigma$) of SnTe bilayer along armchair and zigzag directions are presented in Figures 7a and 7b, respectively, for electrons and holes at 300 K, 500 K and 700 K. There is no clear monotonic change in the value of $S^2\sigma$ with temperature as a result of trade-off between $S$ and $\sigma$ values. The $S^2\sigma$ value remains as high as 40 up to 96.65 mWK$^{-2}$m$^{-1}$ in a wide range of temperatures at a carrier concentration range of the order $10^{12}$ - $10^{13}$ cm$^{-2}$, where the peak is relatively higher for the holes than for the electrons due to the higher electrical conductivity of holes.

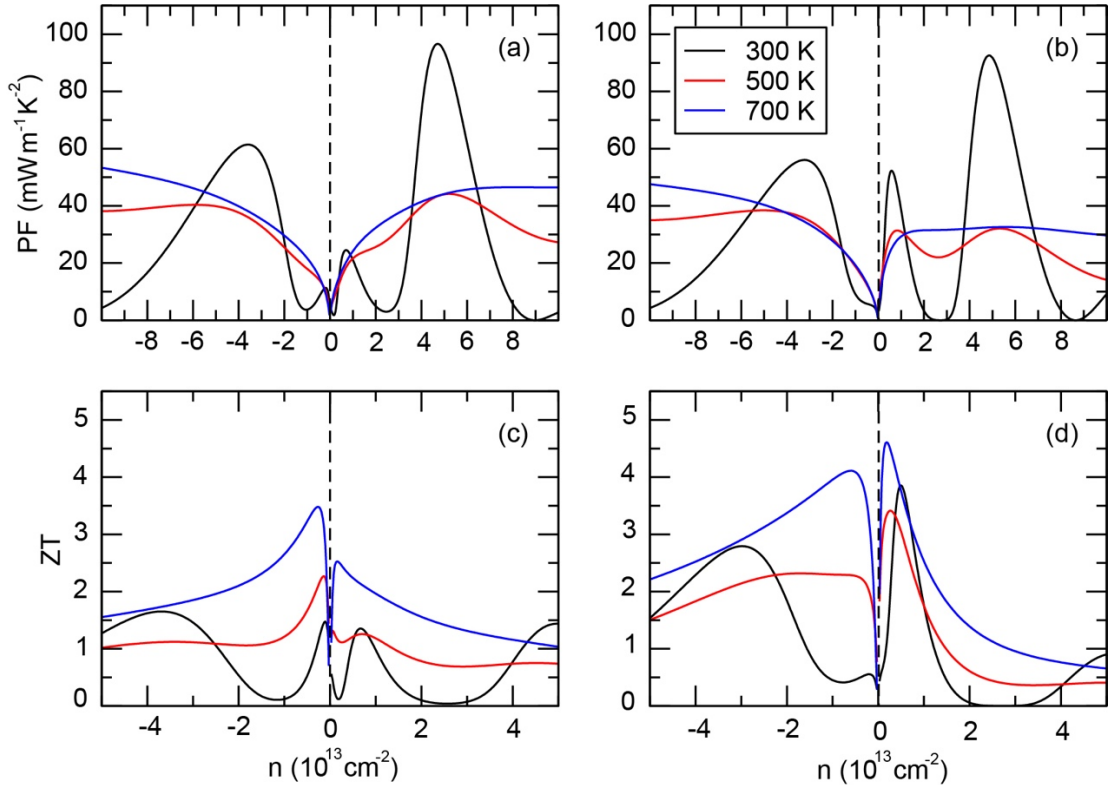

**Figure 7.** (a - b) Power factors and (c - d) TE figure of merit (*ZT*) of AA + s stacked SnTe bilayer as function of carrier concentration along armchair (a, c) and zigzag (b, d) directions.

**Table 2.** Maximum values of figure merit for SnTe bilayer and other phases/compound.

| Compound | Maximum ZT ||
| --- | --- | --- |
|  | *p*-type | *n*-type |
| SnTe bilayer | 4.61 | 4.11 |

| Previous study | SnTe bulk[48] | ~ 0.3 | - |
| | SnTe ML[51] | ~ 2.2 | 2.9 |
| | SnSe bilayer[20] | ~ 0.38 | 0.78 |

The TE figure of merit ($ZT$) is obtained using the calculated TE parameters along armchair and zigzag directions as a function of carrier concentration at different temperatures as shown in Figures 7c and 7d respectively. The value of $ZT$ first increases to reach its peak and then decreases with the carrier concentration in general, which is related to the complicated interdependent behavior of the TE coefficients (S, $\sigma$, $\kappa_e$, and $\kappa_l$) with the carrier concentration. As an example: generally S decreases but $\sigma$ increases with the increase of the carrier concentration (doping). The TE figure of merit ($ZT$) of SnTe bilayer is found to be higher along the armchair direction in the case of *n*-type, whereas the *p*-type shows a higher $ZT$ value along the zigzag direction. Overall, SnTe bilayer shows better TE performance along the zigzag direction than the armchair direction. The predicted $ZT$ peaks are 3.48, 2.27, 1.48 along the armchair and 4.61, 3.41, 3.86 along the zigzag directions at 700K, 500K and 300K, respectively are predicted. These values are significantly higher than those of bulk SnTe,[48] other 2D group IV MLs[15,16,21,38] and SnSe bilayer[20] as compiled in Table 2. The high values of $ZT$ suggest AA + s stacked SnTe bilayer as a promising material for TE device applications and fabrications.

## 4. CONCLUSION

Motivated by the ultralow thermal conductivity and high TE figure of merit of 2D tin chalcogenides, we performed first principles calculations combined with the Boltzmann transport theory to investigate the lattice dynamics, structural, electronic and thermoelectric properties of 2D SnTe bilayer. Structural optimization followed by phonon calculations predicted AA + s stacked SnTe bilayer as the energetically most stable configuration among several stackings. Being a narrow band gap semiconductor, SnTe bilayer is found to have a high Seebeck coefficient, high electrical conductivity and low lattice thermal conductivity that lead to a significantly high ZT value of 4.61 along the zigzag direction. This study not only presents the detailed promising TE properties of SnTe bilayer, but also stimulates further experimental and theoretical studies on few layers of group IV chalcogenides regarding TE performance and device applications in emerging technologies.

AUTHOR INFORMATION

**Corresponding Author**

**Abhiyan Pandit** - *Physics Department, University of Arkansas, Fayetteville, AR 72701, USA;* ORCID iD: https://orcid.org/0000-0003-2158-4597; Email: apandit@uark.edu

**Authors**

**Raad Haleoot** - *Department of Physics at the College of Education, University of Mustansiriyah, Baghdad 10052, Iraq*

**Bothina Hamad** - *Physics Department, University of Arkansas, Fayetteville, AR 72701, USA; Physics Department, The University of Jordan, Amman-11942, Jordan*

**Author Contributions**

A. Pandit performed all the calculations and wrote the manuscript. B. Hamad supervised the work and edited the manuscript. R. Haleoot assisted A. Pandit in the computational tasks of this work.

**Notes**

The authors declare no conflict of interest.

ACKNOWLEDGEMENTS

A. Pandit thanks Dr. T. P. Kaloni for the fruitful discussions. All calculations were performed through Arkansas High Performance Computing Center at the University of Arkansas.